\documentclass[letterpaper, 10 pt, conference]{cssconf}  

\IEEEoverridecommandlockouts                              
\usepackage{amsfonts,amssymb,amsmath} 
\usepackage{sgame}
\usepackage{upgreek}
\usepackage{bm} 
\usepackage[linesnumbered, ruled,noend]{algorithm2e}
\usepackage{graphicx}
\usepackage{mathrsfs}  %
\usepackage{dutchcal} %
\usepackage{tikz}
\usepackage{cleveref}  %
\allowdisplaybreaks %

\newcommand{\nn}{\mathbb{N}}

\newcommand{\EE}{\mathcal{E}}
\newcommand{\FF}{\mathcal{F}}

\renewcommand{\SS}{\mathcal{S}}

\usepackage{xcolor}

\definecolor{ForestGreen}{RGB}{34,139,34}

\newtheorem{assumption}{Assumption} 

\newtheorem{definition}{Definition} 
\newtheorem{lemma}{Lemma} 
\newtheorem{remark}{Remark} 
\newtheorem{theorem}{Theorem} 

\title{\LARGE \bf
Fictitious Play in Extensive-Form Games of Imperfect Information
}

\author{Jason Castiglione and G\"{u}rdal Arslan
	\thanks{G. Arslan is with the Department of Electrical Engineering, University of Hawaii at Manoa,
		Honolulu, HI 96822, USA
		{\tt\small gurdal@hawaii.edu }}%
	\thanks{J. Castiglione is with the Department of Electrical Engineering, University of Hawaii at Manoa,
		Honolulu, HI 96822, USA
		{\tt\small jcastig@hawaii.edu}}%
}

\begin{document}

\maketitle
\thispagestyle{empty}
\pagestyle{empty}

\begin{abstract}

We study the long-term behavior of the fictitious play process in repeated extensive-form games of imperfect information with perfect recall. Each player maintains incorrect beliefs that the moves at all information sets, except the one at which the player is about to make a move, are made according to fixed random strategies, independently across all information sets.  Accordingly, each player makes his moves at any of his information sets to maximize his expected payoff assuming that, at any other information set, the moves  are made according to the empirical frequencies of the past moves. We extend the well-known Monderer-Shapley result \cite{monderer1996fictitious} on the convergence of the empirical frequencies to the set of Nash equilibria to a certain class of extensive-form games with identical interests. We then strengthen this result by the use of inertia and fading memory, and prove the convergence of the realized play-paths to an essentially pure Nash equilibrium in all extensive-form games of imperfect information with identical interests.

\end{abstract}

\section{INTRODUCTION}

The so-called \textit{Fictitious Play (FP)} process
is originally suggested by \cite{bro} as an iterative method for obtaining the value of a zero-sum game,  and its validity is first shown by \cite{robinson1951iterative}. FP and its variants are later adopted as learning processes for multi-player games whereby players myopically seek to maximize their payoffs in a repeated play; see the books \cite{fudlev,young2004strategic} and the references therein. A player employing FP maintains the incorrect belief that his opponents are employing fixed random strategies that can be learned by tracking the empirical frequencies of the past moves. 

FP has been extensively studied principally in the context of finite normal-form games. Much of the existing work is devoted to identifying classes of games in which the beliefs, i.e., the empirical frequencies, generated by FP converge to Nash equilibria. Among other games, FP is shown to be convergent in zero-sum games  \cite{robinson1951iterative}, $2\times2$ games \cite{miyazawa1961convergence},  $2\times n$ games \cite{berger2005fictitious}, identical interest games \cite{monderer1996fictitious}, dominance solvable games \cite{MILGROM199182}, games with ordinal complementarities and diminishing returns \cite{berger2008learning}. 
FP need not converge in all games however. Examples of games in which FP fails to converge in a robust manner appeared in \cite{shapley1963some,jordan1993three,foster1998nonconvergence}. 

Convergence, or lack thereof, in stochastic versions of FP in which randomization is incorporated into player-decision-making is also extensively studied in the literature; see for example \cite{fudenberg1993learning,hofbauer2002global,benaim1999mixed,hofbauer2005learning}. The significance of randomization in player-decision-making and a very appealing robustness result ensuring a near optimal performance for a player using a stochastic FP algorithm (called cautious FP) regardless of the algorithms used by the other players are presented in \cite{cautiousFP}.

On the other hand, far less attention has been paid to the learning processes, in particular FP, in the context of extensive-form games; see our recent work \cite{castiglione2023logit} on valuation-based learning in extensive-form games and the references therein.
The crucial issue in learning in extensive-form games is that player behavior off the realized play-paths are not observed.
 With regards to FP in extensive-form games, we cite \cite{fudenberg1995learning,HENDON1996177} as the relevant references. 

The reference \cite{fudenberg1995learning} considers an FP-like process
for extensive-form games  and show
that, because players observe only the realized play-paths,   
they may fail to learn the behavior of their opponents at information sets that are not 
visited infinitely often. As a result, it is concluded in  \cite{fudenberg1995learning} that, without experimentation, a FP-like process may lead to non-Nash outcomes in general extensive-form games.

The reference \cite{HENDON1996177} considers two versions of FP for preplay reasoning preceding  the actual play of a single round of an extensive-form game. Both versions of FP in \cite{HENDON1996177} require 
players to know the payoff functions of all players to compute   best replies for all players at all information sets  in all rounds, as opposed to observing the best replies of the other players along the  realized play-paths. As such, neither version of FP in \cite{HENDON1996177} is a learning process by which a player can make moves in an actual play with the observations of the realized play paths only, and without the knowledge of the payoff functions or the strategies of his opponents.

In this work, we consider a FP process, for extensive-form games, that is learning process in which players know the game tree and their own payoff functions and observe the realized play-path in each round. Players do not know the payoff functions or the strategies of their opponents and do not observe how their opponents would have moved at the information sets that are not visited in any round. We adopt a version of FP that is based on local best replies, i.e., a player moves at any of his information sets that is reached during the play to maximize his expected payoff in the stage-game by assuming that, at all other information sets including his known, the moves would be made according to the empirical frequencies of the past moves, independently across all information sets.

In an alternative version of FP, a player would make his moves in any round according to a best reply strategy for the stage-game (i.e., by coordinating his moves at all of his information sets) assuming that, at all information sets of the other players, the moves would be made according to the empirical frequencies of the past moves, independently across all information sets.
In one-player-one-move games where a player can move only once \cite{fudenberg1995learning,HENDON1996177}, these two versions of FP lead to the same dynamics; however, they may generate different play-paths in general. 
One justification for the version of FP based on local best replies is that it is simpler for a player to 
compute his best-reply moves at his information sets independently, compared to computing a best-reply strategy for the stage-game, in particular, in games where a player can move many times.

We present two main results. Our first main result shows the convergence of the empirical frequencies to the set of Nash equilibria in  extensive-form games with identical interests under the assumption that the optimality gap at any information set, calculated based on the empirical frequencies, does not increase in any round in which the information set is not reached. We then identify a class of extensive-form games, which includes all games of perfect information as well as all games of almost perfect information, 
in which the assumption of non-increasing optimality gaps is automatically satisfied.
Hence, our first main result extends, in particular, the well-known Monderer-Shapley result \cite{monderer1996fictitious} to a class of extensive-form games which properly contains the normal-form identical-interest games considered in \cite{monderer1996fictitious}. 

Our second main result shows the convergence of the play-paths in all extensive-form games with identical interests, a stronger notion than the convergence of the player beliefs, generated by FP when the players use inertia and fading-memory \cite{marden2009joint}. Such players use fixed step-sizes to update the empirical frequencies at all information sets, thereby emphasizing the more recent observations, and repeat their previous moves at their information sets reached during each round with some fixed probabilities. Our second main result ascertains that, with probability one, the play will eventually settle at some final path that is repeated indefinitely. As a result,  the empirical frequencies for the moves on the final play-path converge to one, and the empirical frequencies converge to a Nash-equilibrium strategy for the stage-game under which the final-path is taken with probability one. 

To the best knowledge of the authors, the main results presented in this paper constitute the first strong convergence results for the FP process in a certain class of extensive-form games. Our model is introduced in Section~\ref{sec:setup}. The details of our generalization of the Monderer-Shapley result \cite{monderer1996fictitious} to extensive-form games are provided in Section~\ref{sec:main}. The details of our stronger convergence results by the use of inertia and fading memory are provided in Section~\ref{sec:fpwifm}. 
Section~\ref{sec:concl} is devoted to conclusions and remarks on future research.

\section{Setup}
\label{sec:setup}

Our general framework is that of extensive-form games of imperfect information. We consider only games of perfect recall, i.e., no player forgets any information that he once possessed. In addition, for simplicity, our model does not include moves by the nature. A precise model is provided below.

\subsection{Stage Game}
The stage game, denoted by $G$, has a finite set $I=\{1,\dots,|I|\}$ of strategic players. 
The game tree consists of a finite set of nodes, which are partially ordered by a precedence relation. There is a single root node $r$, where the stage game starts, which precedees every other node. Each node, except the root node $r$, has a single immediate predecessor. As a result, for any node $n$, there is a unique path from $r$ to $n$, that is a unique set of nodes $(n_1, n_2,\dots, n_{\ell})$, where $n_1=r$, $n_{\ell}=n$, and $n_i$ is the immediate predecessor of $n_{i+1}$ for $i\in[1,\ell)$. A node that does not precede any other node is called a terminal node, and a non-terminal node is called a decision node. We denote the set of terminal nodes by $Z$, and the set of decision nodes by $N$.

A single player moves at any decision node $n\in N$.
The set of nodes in which player $i\in I$ makes moves is denoted by $N^i$, where $N^i \subset N$ and $\cup_{i\in I} N^i = N$.
The set of moves at a decision node $n\in N^i$, available to player $i$, is in one-to-one correspondence with the set of immediate successors of $n$, and is denoted by $A^n$.
Starting from the root node, at each decision node $n\in N^i$ reached during the play of the stage game, player $i$ makes a move $a\in A^n$, which determines the next node reached. The stage game ends when a terminal node $z$ is reached, at which point each player $i\in I$ receives a payoff denoted by $u^i(z)\in\mathbb{R}$.

The set of decision nodes $N$ is partitioned as a set $H$ of information sets such that each node $n\in N$ belongs to a single information set $h\in H$ and $\cup_{h\in H} = N$; the information set containing the root node $r$ is the singleton $\{r\}$.
It is assumed that, for each information set $h\in H$, the same player moves at every node $n\in h$, i.e., $h\subset N^i$ for some $i\in I$.
It is further assumed that, for each information set $h\in H$, the same set $A^n$ of moves are available at every node $n\in h$, which can also be denoted by $A^h$. 
For each $i\in I$, we denote the set of information sets at which player $i$ moves by $H^i$. We note that $\cup_{h\in H^i}=N^i$.
The interpretation is that a player $i$ cannot distinguish between the nodes in any information set $h\in H^i$. We denote by $h(n)$ the information set that contains the node $n\in N$.
In addition, if the path from $r$ to $n\in N\cup Z$ goes through $h\in H$, we write $h\sqsubset  n$; otherwise, we write $h\not\sqsubset n$.  Similarly,  if the path from $r$ to $n\in N\cup Z$ goes through $h\in H$ and  $a\in A^h$, we write $(h,a)\sqsubset  n$.

Each player $i\in I$ makes his/her moves using a strategy $f^i$ defined on $H^i$ such that, for each $h\in H^i$, $f^i(\cdot|h)$ is a probability distribution over the set $A^h$ of moves available at $h$.
If a node $n\in h$ in an information set  $h\in H^i$ is reached during the play, player $i\in I$ makes his move $a\in A^h$ at $n$ according to $f^i(\cdot|h)$ where $a$ is an independent draw from $f^i(\cdot|h)$. For $h\in H$, we sometimes write $f(\cdot|h)$ to mean $f^i(\cdot|h)$ for the player $i\in H$ such that $h\in H^i$.

We use the notation $f:=(f^j)_{j\in I}$ to denote the joint strategy employed by all players, and $f^{-i}:=(f^j)_{j\in I \backslash\{i\}}$ to denote the joint strategy employed by all players except a player $i\in I$. In general, we refer to the set of players other than player $i\in I$ as $-i$, i.e., $-i$ refers to $I\backslash\{i\}$.
We sometimes write $f$ as $f=(f^i,f^{-i})$ for any $i \in I$. 
We denote the set of probability distributions over $A^h$ by $\Delta^h$, the set of strategies for each player $i$ by  $F^i:=\times_{h\in H^i} \Delta^h$, and the set of joint strategies by $F:=\times_{i\in I} F^i$.

For any $h\in H$, we denote the vertex of $\Delta^h$ corresponding to $a\in A^h$ by $\tilde{f}^h_a$, that is the element of $\Delta^h$ assigning probability one to $a$.
For any $h\in H$, $a\in A^h$, $f\in F$, we let $(a,f^{-h})$ denote the joint strategy obtained from $f$ by replacing $f(\cdot|h)$ with $\tilde{f}^h_a$.
Under $(a,f^{-h})$, if $h$ is reached, the move $a$ is taken at $h$  with probability one.
Similarly, for any  $z\in Z$, $f\in F$, we let $(z,f^{-z})$ denote the joint strategy obtained from $f$ by replacing $f(\cdot|h)$ with $\tilde{f}^h_a$ for each $(h,a)\sqsubset z$. Clearly, $z\in Z$ is reached with probability one under $(z,f^{-z})$. We call $f\in F$ a pure (joint) strategy if, for every $h\in H$, $f(\cdot|h)$ assigns probability one to some move $a^h\in A^h$, in which case we denote $f$ also by $a=(a^h)_{h\in H}$ by a slight abuse of notation.

For any node $n\in N \cup Z$, we let $P_{f}[n]$ denote the probability of reaching $n$ under $f\in F$, that is given by 
$$P_f[n]=\prod_{(h,a)\sqsubset n} f(a|h)=f(n_2|h(n_1))\times\cdots\times f(n_{\ell}|h(n_{\ell-1}))$$
where $(n_1,\dots,n_{\ell})$ is the unique path from $r$ to $n$. 
Each player $i$'s expected payoff under a strategy $f\in F$ is given by
$$u^i(f):=E_{f}[u^i]=\sum_{z\in Z} u^i(z)P_{f}[z]. $$

\begin{definition}
	A joint strategy $\bar{f}\in F$ is called a \textit{Nash equilibrium} of the stage game $G$ if
	$$u^i(\bar{f}) = \max_{f^i\in F^i}  u^i(f^i,\bar{f}^{-i}), \qquad \forall i\in I. $$
	We denote the set of Nash equilibria by $\EE$.
\end{definition}

\begin{definition}
	The optimality gap for a joint strategy $f \in F$ at an information set $h\subset H$ is defined by
	$$V_f(h) :=   \max_{a\in A^h} u^i(a,f^{-h}) - u^i(f)$$
	where  $i$ is the player index for which $h\in H^i$. We denote the set of joint strategies optimal at $h\in H$ by $\mathcal{O}(h)$, where
	$$\mathcal{O}(h):=\{f\in F: V_f(h)=0\}.$$
\end{definition}

We observe that 
\begin{align}
	\nonumber V_f(h) & = \max_{a\in A^h}\sum_{z\in Z} u^i(z) P_{(a,f^{-h})}[z] -  \sum_{z\in Z} u^i(z) P_f[z] \\ \label{eq:Vfhalt}
	& = \max_{a\in A^h}\sum_{z\in Z(h)} u^i(z) P_{(a,f^{-h})}[z] -  \sum_{z\in Z(h)} u^i(z) P_f[z] 
\end{align}
where $Z(h)$ denotes the subset of the terminal nodes reachable from $h$. For any $h\in H$, let $P_f[h]$ denote the probability of reaching $h$ under $f$, or $(a,f^{-h})$ for some $a\in A^h$, i.e., $$P_f[h]=\sum_{n\in h} P_f[n] = \sum_{n\in h} P_{(a,f^{-h})}[n] =P_{(a,f^{-h})}[h].$$ If $P_f[h]>0$, then we can write $V_f(h)$ as
\begin{align}
	\nonumber V_f(h) 
	& =P_f[h] \left(\max_{a\in A^h} E_{(a,f^{-h})}[u^i|h] -  E_f[u^i|h] \right).
\end{align}
Therefore, if $P_f[h]>0$, 
\begin{align*}
	f \in \mathcal{O}(h) \qquad\Leftrightarrow\qquad
	E_f[u^i|h] = \max_{a\in A^h} E_{(a,f^{-h})}[u^i|h]. 
\end{align*}
If $P_f[h]=0$, we trivially have $V_f(h)=0$ and $f\in\mathcal{O}(h)$.

Finally, it is clear that 
$$\EE \subset \cap_{h\in H}  \mathcal{O}(h).$$
It turns out that, due to the one-shot-deviation-principle in games of perfect recall \cite{hendon1996one}, we have
$$ \EE=\cap_{h\in H} \mathcal{O}(h) .$$

For future reference, we introduce the set of moves at $h\in H$ that are optimal against $f^{-h}$ as
$$BR^h(f) = \{a\in A^h: (a,f^{-h}) \in \mathcal{O}(h)\}.$$ 

\subsection{Repeated Game and Fictitious Play}
We now introduce a repeated game in which the stage game $G$ is repeated over an infinite number of rounds $t\in\nn$.
The terminal node reached in round $t\in\nn$ is denoted by $z_t \in Z$.
At the end of each round $t\in\nn$, the history of the play is $(z_1,\dots,z_t)$.
The strategy $f_t^i$ used by player $i\in I$ in round $t\in\nn$ can be chosen on the basis of $(z_1,\dots,z_{t-1})$ (the initial history at $t=0$ is the null history). 

We now introduce the fictitious play (FP) process in which players choose their strategies $f_t$ for round $t$ of the repeated game on the basis of 
the empirical frequencies of the past moves of all players at all information sets. Accordingly, for each $h\in H$, we let $f_t(\cdot | h) \in \Delta^h$ denote the empirical frequency of the actions in rounds prior to round $t$ in which the information set $h$ is visited. These empirical frequencies evolve according to: for all $t \in\nn$, $h\in H$, $a \in A^h$,
\begin{align*}
	f_{t+1}(a | h) & = f_t(a | h) + \delta_t(a | h)
\end{align*}
where 
\begin{align}
	\delta_t(a | h) & := \frac{I_t(h)}{\#_t(h)}(I_t(a)-f_t(a | h)) \label{eq:del}\\
	I_t(h) & := \left\{ \begin{array}{cc} 1 & \mbox{if $h$ is visited in round $t$} \\ 0 & \mbox{else} \end{array}\right. \nonumber \\
	I_t(a) & := \left\{ \begin{array}{cc} 1 & \mbox{if $a$ is chosen in round $t$} \\ 0 & \mbox{else} \end{array}\right.  \nonumber \\
	\#_t(h) & := \sum_{k=1}^t I_k(h)  \nonumber 
\end{align}
and $f_1(\cdot|h)$ is arbitrary.
In FP, each player $i$ reaching an information set $h\in H^i$ in round $t$ chooses his move at $h$ to maximize his expected utility with the (incorrect) belief that all other moves (including his own) would be selected according to $f_t=(f_t(\cdot|h))_{h\in H}$. In other words, for any $h\in H$, $a\in A^h$,
\begin{gather}
	I_t(h)I_t(a) =1  \qquad  \Rightarrow  \qquad a\in BR^h(f_t) \label{eq:eqopt} 
\end{gather}

In the next section, we present our main result on the long term behavior of the empirical frequencies  generated by FP.

\section{Convergence of FP in Identical Interest Games}
\label{sec:main}

Or main result concerns the convergence of the empirical frequencies generated by FP in games of identical interests.
\\

\begin{assumption}
	\label{as:1}
	The empirical frequencies $\{f_t\}_{t\in\nn}$ generated by FP satisfies: for all $h\in H$, $t\in\nn$,
	\begin{equation}
		\label{eq:Vnoninc}
		I_t(h)=0 \quad\Rightarrow\quad V_{f_{t+1}}(h) \leq V_{f_t}(h) 
	\end{equation}
	i.e., the optimality gap for $f_t$ at any information set $h\in H$ does not increase in any round $t\in\nn$ in which $h$ is not visited.\\
\end{assumption}

\begin{lemma}
	\label{lem:1}
	Assumption~\ref{as:1} holds in the  following class of games: all immediate successors of all nodes in an information set of size two or higher   belong to  a single information set or they are all terminal nodes.\\
\end{lemma}

\textit{Proof:}
Fix $h\in H$, $z\in Z(h)$, and let $(n_1,\dots,n_{\ell_z})$ be the unique path from $r$ to $z$, where $n_1=r$, $n_{\ell_z}=z$, and $n_m\in h$ for some $m\in[1,\ell_z)$. Assume that, in round $t$, players followed a path 
$(n_1,\dots,n_s,\tilde{n}_{s+1},\dots,)$ that did not reach $h$ and $\tilde{n}_{s+1}\not=n_{s+1}$ for some $s\in [1,m)$. We must have that $h(n_i)=\{n_i\}$, for $i\in[1,s]$, in the class of games considered in this lemma. We have
\begin{align*}
	P_{f_{t+1}}[z] & = \prod_{\ell\in[1,\ell_z)} f_{t+1}(n_{\ell+1}|h(n_{\ell})) \\
	& = \prod_{\ell\in[1,s]}  \frac{\#_{t+1}(n_{\ell+1})}{\#_{t+1}(n_{\ell})}  \prod_{\ell\in[s+1,\ell_z)} \frac{\#_{t+1}(n_{\ell+1})}{\#_{t+1}(h(n_{\ell}))}   \\
	& = \frac{\#_{t+1}(n_{s+1})}{\#_{t+1}(n_1)}  \prod_{\ell\in[s+1,\ell_z)} \frac{\#_t(n_{\ell+1})}{\#_t(h(n_{\ell}))}   \\
	& = \frac{\#_t(n_{s+1})}{t+1}  \prod_{\ell\in[s+1,\ell_z)} f_t(n_{\ell+1}|h(n_{\ell}))   \\
	& =\frac{t}{t+1} 	P_{f_t}[z].
\end{align*}
By the same reasoning, if $h$ is not visited in round $t$, we have, for all $z\in Z(h)$, $a\in A^h$,
\begin{align*}
	P_{\big(a,f_{t+1}^{-h}\big)}[z] 
	& =\frac{t}{t+1} 	P_{\big(a,f_t^{-h}\big)}[z]. 
\end{align*}
By (\ref{eq:Vfhalt}), this results in
$$I_t(h)=0 \qquad\Rightarrow\qquad V_{f_{t+1}}(h) = \frac{t}{t+1} V_{f_t}(h).$$

\begin{remark}
	Figure~3 and Figure~8 in \cite{HENDON1996177} show  examples of games within the class considered in Lemma~\ref{lem:1}.
	Figure~7 in \cite{HENDON1996177} shows an example of a game that is outside the class considered in Lemma~\ref{lem:1}; however, it is straightforward to see that Assumption~\ref{as:1} holds in this game as well. A better characterization of the class of games for which Assumption~\ref{as:1} holds is currently under study.\\
\end{remark}

\begin{theorem}
	\label{th:1}
	Consider a stage game $G$ of perfect recall where players have a common  payoff function $u:Z\to\mathbb{R}$, i.e., $u^i \equiv u$ for all $i\in I$. If Assumption~\ref{as:1} holds, the empirical frequencies  $\{f_t\}_{t\in\nn}$ generated by FP in a repeated play of $G$  converges to the set $\EE$ of equilibria of $G$.\\
\end{theorem}

\begin{proof} 
	The proof follows along similar lines as the proof of Theorem~A in \cite{monderer1996fictitious}. Consider any infinite play-path that can be generated by FP, and let $\{f_t\}_{t\in\nn}$ denote the corresponding sequence of empirical frequencies. 
	Let $H_{\infty} \subset H$ denote the subset of information sets visited infinitely many times, i.e.,
	\begin{align*}
		H_{\infty} & := \{h\in H: \lim_{t\rightarrow\infty} \#_t(h)=\infty\}. 
	\end{align*}
	
	Claim~1: $\lim_{t\rightarrow\infty}V_{f_t}(h)=0$ for all $h\in H\backslash H_{\infty}$. Claim~1 follows from
	$$\lim_{t\rightarrow\infty} P_{f_t}[n]=0, \qquad\forall n\in h, h\in H\backslash H_{\infty}.$$
	To see this, fix $h\in H\backslash H_{\infty}$ and $n\in h$, and let $(n_1,\dots,n_{\ell})$ be the unique path from $r$ to $n$. Since $r$ is visited in every round and $n$ is visited finitely many times, there must be a node $n_m$, $m\in[1,\ell)$, such that $n_m$ is visited infinitely often but $n_{m+1}$ is visted finitely many times. This implies  $\lim_{t\rightarrow\infty} f_t(n_{m+1}|h(n_m))=0$, which in turn implies $\lim_{t\rightarrow\infty} P_{f_t}[n]=0$.
	
	Claim~2:
	  $\lim_{t\rightarrow\infty}V_{f_t}(h)=0$ for all $h\in H_{\infty}$. We have
	  
	\begin{align}
		u(f_{t+1})  - u(f_t)  & =   \sum_{z\in Z} u(z) \Bigg[ \prod_{\ell\in[1,\ell_z)} f_{t+1} (n_{\ell+1} | h(n_{\ell}))   \nonumber	\\ 
		&  \qquad\qquad\qquad- \prod_{\ell\in[1,\ell_z)} f_t (n_{\ell+1} | h(n_{\ell})) \Bigg]  \nonumber	 \\
		&  = \sum_{z\in Z} u(z) \sum_{J \in \mathcal{J}_z}  \Bigg[ \prod_{\ell\in J}   \delta_t (n_{\ell+1} | h(n_{\ell}))    \nonumber	\\ & \qquad \quad  . \times \prod_{\ell\in[1,\ell_z)\backslash J} f_t (n_{\ell+1} | h(n_{\ell}))   \Bigg] 		  \label{eq:Ediff}
	\end{align}
	where 
	$(n_1,\dots,n_{\ell_z})$ denotes the unique path from the root node $r$ to $z\in Z$\footnote{$n_1=r$, $n_{\ell_z}=z$; otherwise, the dependence of $(n_1,\dots,n_{\ell_z})$ on $z$ is suppressed.} , and
	$\mathcal{J}_z$ denotes the set of non-empty subsets of $[1,\ell_z)$.
	We write (\ref{eq:Ediff}) as
	\begin{align}
		u(f_{t+1})  - u(f_t)  = \Delta_t  + \epsilon_t \label{eq:Eftdiff}
	\end{align}
	where 
	\begin{align}
		\Delta_t := & \sum_{z\in Z} u(z)  \sum_{\bar{\ell}\in[1,\ell_z)} \Bigg[   \delta_t (n_{\bar{\ell}+1} | h(n_{\bar{\ell}})) \nonumber \\ &  \qquad \qquad\qquad\quad \times  \prod_{\ell\in[1,\ell_z)\backslash\{\bar{\ell}\}} f_t (n_{\ell+1} | h(n_{\ell}))   \Bigg] \label{eq:diff2}  \\
		\epsilon_t	:=	&  \sum_{z\in Z} u(z) \sum_{J \in \mathcal{J}_z:|J|\geq2}  \Bigg[ \prod_{\ell\in J}   \delta_t (n_{\ell+1} | h(n_{\ell}))  \nonumber \\ & \qquad\qquad\qquad\qquad\quad \times  \prod_{\ell\in[1,\ell_z)\backslash J} f_t (n_{\ell+1} | h(n_{\ell}))   \Bigg]. \nonumber 
	\end{align}
	Due to (\ref{eq:del}), we have 
	\begin{equation}
		\sum_{t\in\nn} \epsilon_t < \infty. \label{eq:epsum}
	\end{equation}
	We rewrite (\ref{eq:diff2}) as
	\begin{align*}
		\Delta_t = & \sum_{h\in\mathcal{H}}  \sum_{a\in A^h}  \sum_{z\in Z(h,a)} \Bigg[u(z)  \delta_t (a | h) \\ & \qquad\qquad\qquad\times \prod_{\ell\in[1,\ell_z):h(n_{\ell})\not=h} f_t (n_{\ell+1} | h(n_{\ell})) \Bigg]				  
	\end{align*}
	where $Z(h,a)\subset Z$ denotes the subset of the terminal nodes that can be reached from $h$ via the move $a \in A^h$.
	We observe that, for each $h\in H$, $a\in A^h$, $z\in Z(h,a)$, 
	$$P_{\left(a,f_t^{-h}\right)}[z] = \prod_{\ell\in[1,\ell_z) : h(n_{\ell})\not=h} f_t (n_{\ell+1} | h(n_{\ell}))$$
	which leads to
	\begin{align*}
		\Delta_t 	 =   & \sum_{h\in\mathcal{H}}  \sum_{a\in A^h}  \delta_t (a | h) \sum_{z\in Z(h,a)} \left[u(z)   P_{\left(a,f_t^{-h}\right)}[z] \right]	
		\\ = & \sum_{h\in\mathcal{H}}  \sum_{a\in A^h}  \delta_t (a | h) \sum_{z\in Z(h)} \left[u(z)   P_{\left(a,f_t^{-h}\right)}[z] \right] 
		\\ = & \sum_{h\in\mathcal{H}}  \sum_{a\in A^h}  \delta_t (a | h) \sum_{z\in Z} \left[u(z)   P_{\left(a,f_t^{-h}\right)}[z] \right] 
	\end{align*}
	where 
	we used the fact that $P_{\left(a,f_t^{-h}\right)}[z]=0$ for all $z\in Z(h) \backslash Z(h,a)$ and $P_{\left(a,f_t^{-h}\right)}[z]=P_{f_t}[z]$ for all $z\in Z\backslash Z(h)$.
	By 
	using (\ref{eq:eqopt}), we obtain
	\begin{align*}
		\Delta_t 	 =   & \sum_{h\in\mathcal{H}} \frac{I_t(h)}{\#_t(h)} \Bigg[ \max_{a\in A^h}       \sum_{z\in Z} \left[u(z)   P_{\left(a,f_t^{-h}\right)}[z]
		\right] \\ & \qquad\qquad\qquad\qquad\qquad-
		\sum_{z\in Z} \left[u(z)   P_{f_t}[z]
		\right] \Bigg] \\   
		= & \sum_{h\in\mathcal{H}} \frac{I_t(h)}{\#_t(h)} \left[ \max_{a\in A^h}       E_{\left(a,f_t^{-h}\right)}[u]
		-
		E_{f_t}[u]
		\right] \\   
		= & \sum_{h\in\mathcal{H}}  \frac{I_t(h)}{\#_t(h)}   V_{f_t}(h).  
	\end{align*}
	Due to (\ref{eq:Eftdiff}) and (\ref{eq:epsum}), we have $\sum_{t\in\nn} \Delta_t<\infty$. 
	For all $h\in H_{\infty}$ and $k\in\nn$, let $t_k(h)$ denote the round $t$ in which the information set $h$ is visited the $k$-th time, i.e., $\#_t(h)=k$.

	We have, for all $h\in H_{\infty}$,
	$$\sum_{k\in\nn} \frac{V_{f_{t_k(h)}}(h)}{k} \leq \sum_{t\in\nn} \Delta_t<\infty.$$
	This implies the following: for all $h\in H_{\infty}$,
	$$\lim_{K\rightarrow\infty} \sum_{k\geq K} \frac{V_{f_{t_k(h)}}(h)}{k}=0$$
	and 
	$$\lim_{\bar{K}\rightarrow\infty} \frac{1}{\bar{K}}\sum_{K\in[1,\bar{K}]} \sum_{k\geq K} \frac{V_{f_{t_k(h)}}(h)}{k}=0$$
	and 
	$$\lim_{\bar{K}\rightarrow\infty}  \frac{1}{\bar{K}}  \sum_{k\in[1,\bar{K}]} V_{f_{t_k(h)}}(h) = 0$$
	and
	\begin{equation}
		\lim_{\bar{K}\rightarrow\infty}  \frac{1}{\bar{K}}  \sum_{k\in[1,\bar{K}]} I\{ V_{f_{t_k(h)}}(h) > \epsilon\} = 0, \quad \forall \epsilon>0. \label{eq:cm1}
	\end{equation}
	From (\ref{eq:cm1}), we obtain, for all $h\in H_{\infty}$, $\epsilon>0$,
	$$\lim_{\bar{K}\rightarrow\infty}  \frac{1}{\bar{K}}   \sum_{k\in[1,\bar{K}]} I\{ f_{t_k(h)} \not \in B_{\epsilon}(\mathcal{O}(h))\} = 0
	$$
	where
	$$B_{\epsilon}(\mathcal{O}(h)):=\{f\in F: |f-g|<\epsilon, \ \mbox{for some} \ g\in\mathcal{O}(h)\}.$$
	
	Lemma~1 in \cite{monderer1996fictitious} implies that, as $k\rightarrow\infty$,
	$$f_{t_k(h)} \rightarrow \mathcal{O}(h), \qquad \forall h\in H_{\infty}$$
	which in turn implies
	\begin{equation}
		\label{eq:Vthconv}	
		\lim_{k\rightarrow\infty} V_{f_{t_k(h)}}(h)= 0, \qquad \forall h\in H_{\infty}.
	\end{equation}
	
	We now argue that, under Assumption~\ref{as:1},  (\ref{eq:Vthconv}) implies
	\begin{equation}
		\label{eq:Vtconv}
		\lim_{t\rightarrow\infty} V_{f_t}(h) = 0, \qquad \forall h\in H_{\infty}.
	\end{equation}
	We note that, for every $h\in H$, $f \mapsto V_f(h)$ is continuous and $\lim_{t\rightarrow\infty}|f_{t+1}-f_t|=0$, therefore,
	$$\lim_{t\rightarrow\infty}|V_{f_{t+1}}(h)-V_{f_t}(h)|=0.$$   
	In view of this, (\ref{eq:Vthconv}) implies (\ref{eq:Vtconv}), under Assumption~\ref{as:1}.
	The proof follows from Claim~1 and Claim~2 due the continuity of $f$.
\end{proof}

\begin{remark}
	Theorem~\ref{th:1} holds under the following assumption as well; however, it is not clear how to ensure that this assumption holds.
\end{remark}

\begin{assumption}
	\label{as:2}
	The empirical frequencies $\{f_t\}_{t\in\nn}$ generated by FP satisfies: for all $h\in H_{\infty}$, 
	\begin{equation}
		\lim_{k\rightarrow\infty} \frac{t_{k+1}(h)-t_k(h)}{t_k(h)} =0.
	\end{equation}
\end{assumption}

\section{Fictitious Play with Inertia and Fading Memory}
\label{sec:fpwifm}

Here, we replace the empirical frequency dynamics with: for all $t \in\nn$, $h\in H$, $a \in A^h$,
\begin{align}
	f_{t+1}(a | h) & = f_t(a | h) +\rho I_t(h)(I_t(a)-f_t(a | h)) \label{eq:Fmap}
\end{align}
where $\rho\in(0,1)$ is the fading memory parameter and $f_1\in F$ is arbitrary. In addition, the players employ the following strategies in each round $t\in\nn$: starting with an arbitrary pure (joint) strategy 
$a_1=(a_1^h)_{h\in H}$, 
if $h\in H$ is reached in round $t\geq2$ and $ a_{t-1}^h \notin BR^h(f_t)$, then
$$\left\{\begin{array}{cl}
	a_t^h=a_{t-1}^h, & \textrm{with probability} \ \alpha^h \\
	a_t^h\in BR^h(f_t), & \textrm{with probability} \ 1-\alpha^h 
\end{array} \right.$$
otherwise,  $$a_t^h=a_{t-1}^h$$ where 
$\alpha^h\in(0,1)$ is the inertia parameter for each $h\in H$.

In each round $t\in\nn$, $z_t$ is reached under the pure strategy $a_t=(a_t^h)_{h\in H}$.  
If $h$ is reached during rounds $1,\dots,t$, $a_t^h$ denotes the most recent move at $h$ during rounds $1,\dots,t$; otherwise $a_t^h=a_1^h$.

\begin{theorem}
	\label{th:2}
	Consider a stage game $G$ of perfect recall where players have a common  payoff function $u:Z\to\mathbb{R}$, i.e., $u^i \equiv u$ for all $i\in I$. If $u(z)\not=u(\bar{z})$ for all terminal nodes $z\not=\bar{z}$, then,  with probability one, the empirical frequencies  $\{f_t\}_{t\in\nn}$ generated by FP with inertia and fading memory in a repeated play of $G$  converges to a strategy $f^*$ in the set $\EE$ of equilibria of $G$; moreover,  for some $z^*\in Z$, $f^*(a|h)=1$, for all $(h,a)\sqsubset z^*$. \\
\end{theorem}

We next present several definitions and lemmas that will lead to a proof of Theorem~\ref{th:2}.\\

We let $K_{\max}\in\nn_0$ denote the smallest integer such that
\begin{equation}
	1-(1-\rho)^{K_{\max}+1}\geq \left(1-\frac{\Delta_{\min}}{8u_{\max}}\right)^{1/L_{\max}}.
	\label{eq:Kdef}
\end{equation}
where $L_{\max}$ denotes the maximum number of moves required to traverse any path from $r$ to any terminal node, and
$u_{\max} := \max_{z\in Z}|u(z)|$, $\Delta_{\min}  :=\min_{z\not=\bar{z}} |u(z)-u(\bar{z})|$.
Note that $\Delta_{\min}>0$ by assumption, and $\Delta_{\min}\leq 2 u_{\max}$.
For any $t\in\nn$, we let $K_t\in\nn_0$ be the smallest integer such that
\begin{align}
	& (1-\rho)^{K_t+1} f_t(a|h)+1-(1-\rho)^{K_t+1} \nonumber \\ & \qquad \qquad \geq \left(1-\frac{\Delta_{\min}}{8u_{\max}}\right)^{1/L_{\max}}, \ \forall(h,a)\sqsubset z_t.
	\label{eq:Ktdef}
\end{align}
When $z_t=z_{t+1}=\cdots=z_{t+K_t}$, the left-hand-side of (\ref{eq:Ktdef}) equals $f_{t+K_t+1}(a|h)$. Therefore, $K_t$ represents the minimum number of repetitions of $z_t$  required for the empirical frequencies $f_{t+K_t+1}(a|h)$, $(h,a)\sqsubset z_t$, to satisfy the lower bound on the right-hand-side of  (\ref{eq:Ktdef}). $K_t$ depends on $t$ only through $(f_t,z_t)$,  and the uniform upper-bound $K_t \leq K_{\max}$ holds for all $t\in\nn$, regardless of what $(f_t,z_t)$ is. 
Therefore, for any $t\in\nn$, the ``repeat event'' 
$$E_t:=\{z_t=z_{t+1}=\cdots=z_{t+K_t}\}$$
ensures
\begin{equation}
	f_{t+K_t+1}(a|h)\geq \left(1-\frac{\Delta_{\min}}{8u_{\max}}\right)^{1/L_{\max}}, \ \forall(h,a)\sqsubset z_t.
	\nonumber
\end{equation}

We let $\FF_1:F\times Z \to F$ denote the (deterministic) mapping $(f_t,z_t) \to f_{t+1}$ defined by (\ref{eq:Fmap}), i.e., $f_{t+1}=\FF_1(f_t,z_t)$. 
We then define the mappings $\FF_2,\FF_3,\dots$ by the recursion
$$\FF_{m+1}(f,z)=\FF_1(\FF_m(f,z),z), \quad \forall (f,z)\in F\times Z, m\in\nn.$$
We let $\FF_0(f,z)=f$ for all $(f,z)\in F\times Z$ for notational consistency. 
Note that, for all $t,m\in\nn$, $$\ z_t=z_{t+1}=\cdots=z_{t+m-1} \quad\Rightarrow\quad f_{t+m}=\FF_m(f_t,z_t).$$

\begin{definition}
	For any $\ell\in\nn$, we say that $(f,z)$ is an  $\ell$-step locked state if $$ a\in BR^h(\FF_m(f,z)), \quad \forall (h,a)\sqsubset z, m\in\{1,\dots,\ell\}.$$ 
	We say that $(f,z)$ is a  locked state if $$ a\in BR^h(\FF_m(f,z)), \quad \forall (h,a)\sqsubset z, m\in\nn.$$ 	
	We denote the subset of the $\ell$-step locked states by $\Lambda_{\ell}$ and the subset of  locked states   by $\Lambda_{\infty}$. 
\end{definition}

Clearly, $F\times Z \supset \Lambda_1 \supset \Lambda_2 \supset  \Lambda_3 \supset\cdots \supset\Lambda_{\infty}=\cap_{\ell=1}^{\infty} \Lambda_{\ell}$ and, for all $t\in\nn$, the following implications hold surely:
\begin{align*}
	(f_t,z_t)\in \Lambda_{\ell} & \quad\Rightarrow\quad z_{\tau}=z_t, \ \forall \tau\in\{t+1,\dots,t+\ell\}   \\ (f_t,z_t)\in  \Lambda_{\infty} & \quad\Rightarrow\quad z_{\tau}=z_t, \ \forall \tau\geq t.
\end{align*}
For any $t\in\nn$, we let $m_t\in\nn_0$ denote the smallest integer such that
\begin{equation}
	(\FF_{K_t}(f_t,z_t),z_t) \not \in \Lambda_{m_t+1}
	\label{eq:nLm}
\end{equation}
where we set $m_t:=\infty$ if $	(\FF_{K_t}(f_t,z_t),z_t)  \in \Lambda_{\infty}$. 
In words, $m_t$ represents the number of forced repetitions of $z_t$ following the repeat event $E_t$, and it depends on $t$ only through $(f_t,z_t)$.

For any $t\in\nn$, if $m_t<\infty$, we define the ``repeat-lock event'' as
\begin{align}
	\bar{E}_t  :=& E_t \cap \{z_{t+K_t}=\cdots=z_{t+K_t+m_t} \} \nonumber \\ & =\{z_t=\cdots=z_{t+K_t+m_t}\}.
	\label{eq:extRt}
\end{align}
If $m_t=\infty$, we let $\bar{E}_t:=\{z_{\tau}=z_t, \ \forall \tau\geq t\}$.
We note that the repetitions of $z_t$ in $\bar{E}_t$  following $E_t$ are forced repetitions.  Clearly, for any $t\in\nn$,  the repeat-lock event $\bar{E}_t$  ensures 
\begin{equation}
	f_{t+K_t+m_t+1}(a|h)\geq \left(1-\frac{\Delta_{\min}}{8u_{\max}}\right)^{1/L_{\max}}, \ \forall(h,a)\sqsubset z_t.
	\nonumber
\end{equation}

If $m_t<\infty$, given the event $\bar{E}_t$, we have $(f_{t+K_t+m_t},z_{t+K_t+m_t})\notin\Lambda_1$ 
due to (\ref{eq:nLm}); therefore, there exists an alternative move $\hat{a}\in A^{\bar{h}}$  at some information set $\bar{h}\sqsubset z_{t+K_t+m_t}$  yielding 
\begin{equation}
	u(\hat{a},f_{t+K_t+m_t+1}^{-\bar{h}}) > u(\bar{a},f_{t+K_t+m_t+1}^{-\bar{h}})
	\label{eq:dev}
\end{equation} 
where $\bar{a}\in A^{\bar{h}}$  is such that $(\bar{h},\bar{a})\sqsubset z_{t+K_t+m_t}$. Accordingly, let $\hat{Z}_{t+K_t+m_t+1}$ denote the subset of the terminal nodes that can be reached by deviating from $z_{t+K_t+m_t}$ at some $\bar{h}\sqsubset z_{t+K_t+m_t}$ via an alternative move $\hat{a}\in A^{\bar{h}}$ satisfying (\ref{eq:dev}),
and otherwise repeating the previous moves, i.e., 
$a_{t+K_t+m_t+1}^{\bar{h}}=\hat{a}$ and
$a_{t+K_t+m_t+1}^{h}=a_{t+K_t+m_t}^{h}$ for all $h\in H \backslash \{\bar{h}\}$.
Next, for any $t\in\nn$, if $m_t<\infty$, we define the ``repeat-lock-deviate event'' as
$$\hat{E}_t:=\bar{E}_t \cap \{z_{t+K_t+m_t+1}\in \hat{Z}_{t+K_t+m_t+1}\}$$
which occurs with probability at least
$$
	\min_{h\in H}(1-\alpha^h)\left(\min_{h\in H} \alpha^h\right)^{L_{\max}(K_{\max}+1)}.
$$

\begin{lemma}
	\label{lem:improve}
	Suppose that, for some $t_0\in\nn$ and $n\in\nn_0$, the following event occurs:
	\begin{equation}
		\hat{E}_{t_0} \cap \hat{E}_{t_1}  \cap\dots\cap\hat{E}_{t_n}
		\label{eq:Rtnseq} 
	\end{equation}
	where  $$t_{\ell+1}:=t_{\ell}+K_{t_{\ell}}+m_{t_{\ell}}+1, \qquad \forall \ell\in\{0,\dots,n\}$$ and $m_{t_0},\dots,m_{t_n}<\infty$.
	Then, we have, for each $\ell\in\{0,\dots,n\}$,
	$$z_{t_{\ell+1}} \in \{z_{t_0},\dots,z_{t_{\ell}}\} \qquad \Rightarrow \qquad u(z_{t_{\ell+1}})>u(z_{t_{{\ell}}}).$$  
\end{lemma}

\begin{proof} 
	For each $\ell\in\{0,\dots,n\}$, the event $\hat{E}_{t_{\ell}}$ implies that $u(\hat{a},f_{t_{\ell+1}}^{-\bar{h}}) > u(\bar{a},f_{t_{\ell+1}}^{-\bar{h}})$ for some $(\bar{h},\bar{a})\sqsubset z_{t_{\ell}}$ and $(\bar{h},\hat{a})\sqsubset z_{t_{\ell+1}}$; moreover, 
	$$P_{(\bar{a},f_{t_{\ell+1}}^{-\bar{h}})}[z_{t_{\ell}}] \geq 1-\frac{\Delta_{\min}}{8u_{\max}}.$$
	We note that $z_{t_{\ell}}$ and $z_{t_{\ell+1}}$ follow the same path from the root $r$ to $\bar{h}$, hence, 
	for each $(h,a)\sqsubset z_{t_{\ell+1}}$ upstream from $\bar{h}$, we have
	$$f_{t_{\ell+1}}(h,a)\geq \left(1-\frac{\Delta_{\min}}{8u_{\max}}\right)^{1/L_{\max}}.$$
	Assume that $z_{t_{\ell+1}}$ is reached at least once during rounds $t_0,\dots,t_{\ell}$, i.e., $z_{t_{\ell+1}} \in \{z_{t_0},\dots,z_{t_{\ell}}\}$. Furthermore, $z_{t_{\ell+1}}$  is reached in round $t_{\ell+1}$ by 
	following the most recent moves, prior to $t_{\ell+1}$, at each information set $\tilde{h}$ downstream from $(h,\hat{a})$, i.e., $a_{t_{\ell+1}}^{\tilde{h}}=a_{t_{\ell}}^{\tilde{h}}$.
	Therefore, for each $(\tilde{h},\tilde{a})\sqsubset z_{t_{\ell+1}}$ downstream from $(\bar{h},\hat{a})$, we have
	$$f_{t_{\ell+1}}(\tilde{h},\tilde{a})\geq \left(1-\frac{\Delta_{\min}}{8u_{\max}}\right)^{1/L_{\max}}.$$
	As a result, we have
	$$P_{(\hat{a},f_{t_{\ell+1}}^{-\bar{h}})}[z_{t_{\ell+1}}] \geq 1-\frac{\Delta_{\min}}{8u_{\max}}.$$
	This gives us
	\begin{align*}
		0 < &
		u(\hat{a},f_{t_{\ell+1}}^{-\bar{h}})- u(\bar{a},f_{t_{\ell+1}}^{-\bar{h}})  \\ 
		= &
		\sum_{z\in Z(\bar{h})} u(z) P_{(\hat{a},f_{t_{\ell+1}}^{-\bar{h}})} [z] -\sum_{z\in Z(\bar{h})} u(z) P_{(\bar{a},f_{t_{\ell+1}}^{-\bar{h}})} [z]  \\
		\leq 	& u(z_{t_{\ell+1}}) P_{(\hat{a},f_{t_{\ell+1}}^{-\bar{h}})} [z_{t_{\ell+1}}] - u(z_{t_{\ell}}) P_{(\bar{a},f_{t_{\ell+1}}^{-\bar{h}})} [z_{t_{\ell}}] \\
		& + \left(1-P_{(\hat{a},f_{t_{\ell+1}}^{-\bar{h}})}[z_{t_{\ell+1}}]+1-P_{(\bar{a},f_{t_{\ell+1}}^{-\bar{h}})}[z_{t_{\ell}}]\right) u_{\max}   \\
		\leq 	&\left(  u(z_{t_{\ell+1}}) - u(z_{t_{\ell}})  \right) P_{(\hat{a},f_{t_{\ell+1}}^{-\bar{h}})} [z_{t_{\ell+1}}] \\
		& + \left|  P_{(\hat{a},f_{t_{\ell+1}}^{-\bar{h}})} [z_{t_{\ell+1}}] -  P_{(\bar{a},f_{t_{\ell+1}}^{-\bar{h}})} [z_{t_{\ell}}] \right| u_{\max} 
		+ \frac{\Delta_{\min}}{4}\\
		\leq 	&\left(  u(z_{t_{\ell+1}}) - u(z_{t_{\ell}})  \right) P_{(\hat{a},f_{t_{\ell+1}}^{-\bar{h}})} [z_{t_{\ell+1}}] + \frac{\Delta_{\min}}{8}
		+ \frac{\Delta_{\min}}{4} \\
		\leq 	&u(z_{t_{\ell+1}}) - u(z_{t_{\ell}})  + 2\frac{\Delta_{\min}}{8}+ \frac{\Delta_{\min}}{8}
		+ \frac{\Delta_{\min}}{4} \\
		<	&u(z_{t_{\ell+1}}) - u(z_{t_{\ell}})  + \Delta_{\min}.
	\end{align*}
	This implies that $0<u(z_{t_{\ell+1}}) - u(z_{t_{\ell}})$, i.e., the lemma holds true.
\end{proof}

\begin{lemma}
	\label{lem:nmax}
	The largest integer $n\in\nn_0$ for which the event  $\hat{E}_{t_0} \cap \hat{E}_{t_1}  \cap\dots\cap\hat{E}_{t_n}$ in Lemma~\ref{lem:improve}  can occur, i.e., $\{m_{t_0},\dots,m_{t_n}<\infty\}$ can occur, is upper bounded as $n< (|Z|-1)|Z|$.
\end{lemma}

\begin{proof}
	Suppose that $\hat{E}_{t_0} \cap \hat{E}_{t_1}  \cap\dots\cap\hat{E}_{t_n}$ occurs for some $t_0\in\nn$ and $n\geq (|Z|-1)|Z|$.
	Let
	the set of  terminal nodes reached during rounds $t_0,\dots,t_k$ be denoted by $Z_k$ for each $k\in\{0,\dots,n+1\}$. Clearly, 
	$Z_0 \subseteq Z_1 \subseteq \cdots \subseteq Z_{n+1} \subseteq Z$. 
	Due to Lemma~\ref{lem:improve}, the following implication must be true for any $k\in\{0,\dots,n-|Z|+1\}$:
	\begin{align}
		& Z_k=Z_{k+1}=\cdots=Z_{k+|Z|} \nonumber \\ &  \Rightarrow 
		u(z_{t_k})<u(z_{t_{k+1}})<u(z_{t_{k+2}})<\dots<u(z_{t_{k+|Z|}}).
		\label{eq:impl}
	\end{align} 
	However, the right-hand-side of (\ref{eq:impl}) cannot occur since the number of terminal nodes is $|Z|$. Therefore, for any $k\in\{0,\dots,n-|Z|+1\}$, we have $|Z_{k+|Z|}| \geq |Z_k|+1$. 
	Since $|Z_1|=2$, this implies
	$$|Z_{1+(|Z|-1)|Z|}| \geq |Z_1|+|Z|-1=|Z|+1$$ which contradicts $|Z_{1+(|Z|-1)|Z|}| \leq |Z|$. This implies the lemma.
\end{proof}

\begin{definition}
	For any $t_0\in\nn$, if $m_{t_0}<\infty$, we define the maximal repeat-lock-deviate event as
	$$M_{t_0}:=\hat{E}_{t_0} \cap \hat{E}_{t_1}  \cap\dots\cap\hat{E}_{t_n}\cap E_{t_{n+1}}\cap \{m_{t_{n+1}}=\infty\}$$ 
	where $n<(|Z|-1)|Z|$ and $t_1,\dots,t_{n+1}$ 
	are as defined in Lemma~\ref{lem:improve}; otherwise, if $m_{t_0}=\infty$, we define the maximal repeat-lock-deviate event as
	$M_{t_0}:=E_{t_0}.$ 
\end{definition}

\begin{lemma}
	\label{lem:prmt}
	For any $t\in\nn$,  the conditional probability of the maximal repeat-lock-deviate event satisfies $$\mathrm{Pr}[M_t|\SS_t]\geq p_{\min}$$
	where  $\SS_t:=(f_1,a_1,\dots,f_t,a_t)$ and
	$$ p_{\min}:=\left(\min_{h\in H} (1-\alpha^h)\big(\min_{h\in H} \alpha^h\big)^{L_{\max}(K_{\max}+1)}\right)^{|Z|^2}.$$
	
\end{lemma}

\begin{proof} 
	Fix any $t\in\nn$.
	Due to inertia,   the conditional probability of the repeat event
	$E_{t}$ satisfies
	\begin{equation}
		\mathrm{Pr}[E_{t}|\SS_{t}] \geq \left(\min_{h\in H} \alpha^h\right)^{L_{\max}K_{\max}}.
		\label{eq:PrEt}
	\end{equation}
	If $m_{t}=\infty$, we have
	$\mathrm{Pr}[M_{t}|\SS_{t}] =\mathrm{Pr}[E_{t}|\SS_{t}]$, which implies $\mathrm{Pr}[M_{t}|\SS_{t}]\geq p_{\min}$.

	We therefore consider the case where $m_{t}<\infty$.
	We observe that the forced repetitions $\{z_{t+K_{t}}=\cdots=z_{t+K_{t}+m_{t}} \}$, given $(\SS_{t},E_{t})$, occurs with probability one.  As a result,  we have
	$\mathrm{Pr}[\bar{E}_{t}|\SS_{t}]  =\mathrm{Pr}[E_{t}|\SS_{t}]$.
	Therefore, $\mathrm{Pr}[\bar{E}_{t}|\SS_{t}]$ also satisfies the lower bound on the right-hand-side of (\ref{eq:PrEt}). 
	
	In addition, given $(\SS_{t},\bar{E}_{t})$, the single-deviation event 
	$ \{z_{t+K_{t}+m_{t}+1}\in \hat{Z}_{t+K_{t}+m_{t}+1}\}$
	occurs with probability at least
	$\min_{h\in H}(1-\alpha^h)\left(\min_{h\in H} \alpha^h\right)^{L_{\max}}.$
	This implies 
	\begin{equation}
		\label{eq:Stpmin}
		\mathrm{Pr}[\hat{E}_{t}|\SS_{t}] \geq 
		\min_{h\in H}(1-\alpha^h)\left(\min_{h\in H} \alpha^h\right)^{L_{\max}(K_{\max}+1)}.
	\end{equation}
	Now, for any $t_0\in\nn$, we write
	\begin{align*}
		& \mathrm{Pr}[M_{t_0}|\SS_{t_0}] \\ & \quad =  \mathrm{Pr}[\hat{E}_{t_0}|\SS_{t_0}]\times
		\mathrm{Pr}[ \hat{E}_{t_1}  |\SS_{t_0},\hat{E}_{t_0}]  \\ & \qquad \times \dots \times
		\mathrm{Pr}[ \hat{E}_{t_n}  |\SS_{t_0},\hat{E}_{t_0},\dots,\hat{E}_{t_{n-1}}] \\ & \qquad \times 	\mathrm{Pr}[ E_{t_{n+1}} \cap \{m_{t_{n+1}}<\infty\}  |\SS_{t_0},\hat{E}_{t_0},\dots,\hat{E}_{t_{n}}] 
	\end{align*}
	Lemma~\ref{lem:nmax} and the lower bound in  (\ref{eq:Stpmin}) lead to the desired result.
\end{proof}

\begin{lemma}
	$	\mathrm{Pr}[(f_t,z_t)\in\Lambda_{\infty} \ \textrm{for some} \ t\in\nn|\SS_1]=1.$
\end{lemma}

\begin{proof}
	By Lemma~\ref{lem:prmt}, for all $t\in\nn$, there exists $T_t\in\nn$ such that  
	\begin{equation}
		\mathrm{Pr}[(f_{t+T_t},z_{t+T_t})\in\Lambda_{\infty} |\SS_t] \geq p_{\min}.
		\label{eq:fTt}
	\end{equation}	
	Let $\tau_0:=1$, and $\tau_{k+1}=\tau_k+T_{\tau_k}$ for all $k\in\nn_0$. By (\ref{eq:fTt}), we have, for all $k\in\nn$, 
	\begin{align*}
		& \mathrm{Pr}[\cap_{\ell\in\nn}\{(f_{\tau_\ell},z_{\tau_\ell})\notin\Lambda_{\infty}\} |\SS_1] \\ & \quad \leq  \mathrm{Pr}[\cap_{\ell=1}^k\{(f_{\tau_\ell},z_{\tau_\ell})\notin\Lambda_{\infty}\} |\SS_1] \\ & \quad \leq
		\prod_{\ell=1}^k
		\mathrm{Pr}[(f_{\tau_\ell},z_{\tau_\ell})\notin\Lambda_{\infty} |\SS_1,\cap_{m=1}^{\ell-1}\{(f_{\tau_m},z_{\tau_m})\notin\Lambda_{\infty}\}]  \\ & \quad \leq (1-p_{\min})^k
	\end{align*}
	This results in
	$$ \mathrm{Pr}[\cap_{\ell\in\nn}\{(f_{\tau_\ell},z_{\tau_\ell})\notin\Lambda_{\infty}\} |\SS_1]=0$$
	which proves the lemma. 
\end{proof}

\begin{lemma}
	If $(f_t,z_t)\in \Lambda_{\infty}$, for some $t\in\nn$, then  $$\lim_{\tau\to\infty} f_{\tau} =(z_t,f^{-z_t})$$ where
	$(z_t,f^{-z_t})\in\EE$.
\end{lemma}

\begin{proof} 	Since $z_{\tau}=z_t$ for all $\tau\geq t$, we have 
	$$\lim_{\tau\to\infty} f_{\tau}(a|h)=1, \quad \forall (h,a)\sqsubset z_t$$ and
	$$f_{\tau}(\cdot|h)=f(\cdot|h), \quad \forall h\not\sqsubset z_t, \tau\geq t.$$
	Therefore, we have $\lim_{\tau\to\infty} f_{\tau}=(z_t,f^{-z_t})$.
	
	We  note that
	$f_{t+m}=\FF_m(f_t,z_t)$, for all $m\in\nn$, and therefore 
	$$u(a,f_{t+m}^{-h})-\max_{\hat{a}\in A^h}u(\hat{a},f_{t+m}^{-h})\geq0, \quad \forall (h,a)\sqsubseteq z_t, m\in\nn.$$
	By letting $f_{\infty}:=\lim_{m\to\infty}f_{t+m}=(z_t,f^{-z_t})$ and using the continuity of 
	$u(g)=\sum_{z\in Z} P_g[z]u(z)$ with respect to $g\in F$, we obtain
	$$u(f_{\infty})-\max_{\hat{a}\in A^h}u(\hat{a},f_{\infty}^{-h})\geq0, \quad \forall h\sqsubseteq z_t$$
	i.e., 
	$f_{\infty}\in\mathcal{O}(h)$, for all $h\sqsubseteq z_t.$
	This together with $P_{f_{\infty}}[h]=1$, for all $h\sqsubset z_t$,  implies that $f_{\infty}$ is an equilibrium. 
\end{proof}
\vspace*{2.3mm}
The proof for Theorem~\ref{th:2} is readily extendable to appropriately defined extensive-form \textit{potential games}.

\section{Conclusions and Future Works}
\label{sec:concl}
We presented an FP process as a learning process in extensive-form games of imperfect information. 
We extended the Monderer-Shapley convergence result \cite{monderer1996fictitious} to a certain class of  extensive-form games with identical interests. 
We then obtained a stronger convergence result, namely the convergence of the play-paths, generated by the FP process with inertia and fading memory in all extensive-form games of imperfect information with identical interests.

Extending the Monderer-Shapley convergence result \cite{monderer1996fictitious} to all extensive-form games with identical interests as well as proving the convergence of play-paths  by the use of inertia only (and without fading memory)
are currently under study. Obtaining comparable convergence results for other classes of extensive-form games without identical interests, e.g., two-player zero-sum games, remains as a significant future research topic.


\bibliographystyle{ieeetr}
\bibliography{references}
\nocite{}

\end{document}